# Incommensurate charge and spin density wave order in electron doped $SrMn_{1-x}W_xO_{3-\delta}$ (x= 0.08 to 0.1875)

Poonam Yadav[1], Shivani Sharma [2,4], Aga Shahee[3], Ivan da Silva[4], Vaclav Petricek[5] and N P Lalla[1]

[1] *UGC-DAE Consortium for Scientific research, Indore-452001, India*
[2] *Jawaharlal Nehru Center for Advanced Scientific Research, Bengaluru, India-560064*
[3] *Center for Novel State of Complex Materials Research, Department of Physics and Astronomy, Seoul National University, Seoul 08826, Republic of Korea*
[4] *ISIS Facility, Rutherford Appleton Laboratory, Chilton, Didcot OX11 0QX, United Kingdom*
[5] *Institute of Physics CAS, Na Slovance 1999/2, Praha, Czech Republic*

## Abstract

Incommensurate (IC) charge-order (CO) and spin (SDW) density wave order in electron doped $SrMn_{1-x}W_xO_{3-\delta}$ (x= 0.08 to 0.1875) have been studied using neutron diffraction (NPD). The study highlights the drastic effect of electron doping on the emergence of magnetic ground states which were not revealed in manganites before. With increasing ($x$) the crystal structure changes from simple tetragonal (*P*4*/mmm*) to an IC-CO modulated structure with super space-group $P2/m(\alpha\beta0)00$ having *ab*-planer ferro order of $\mathbf{3}\boldsymbol{d_{x^2-y^2}}$ orbitals in a compressed tetragonal (c<a) lattice. The IC-CO order is found to be intimately related with the $\mathbf{3}\boldsymbol{d_{x^2-y^2}}$ orbital order. The occurrence of IC-CO has been attributed to the mixed character (itinerant/localized) of $e_g$-electrons undergoing Fermi-surface nesting of $\mathbf{3}\boldsymbol{d_{x^2-y^2}}$ band causing electronic instability, which opens a gap through a charge density wave (CDW) mechanism. This feature appears to share proximity with the high-$T_c$ cuprates. At lower temperatures the CDW phase under goes SDW transition, which changes continuously with ($x$) and finally disappear at higher ($x$) due to introduction of large frustration into the system. For $0.08 \leq x \leq 0.10$ a *C*-type antiferromagnetic (AFM) order with propagation vector k = (1/2, 1/2, 0) appears under ferro-ordering of $\boldsymbol{3d_{z^2}}$ orbitals, whereas for $x > 0.1$, a different *C*-type AFM order with propagation vector **k** = (1/2,0,1/2), coexists with an incommensurate SDW order with k = (0.12, 0.38, 1/2). For compositions with $0.1625 \leq x \leq 0.175$, while the structural features of CDW and orbital-order remain qualitatively the same, the magnetic interaction gets modified and results another SDW phase with single incommensurate propagation vector k = (0.07, 0.43, 1/2). The SDW order has been attributed to introduction of frustration in the spin system. A detail magnetic and structural phase-diagram, as a function of W substitution for $SrMn_{1-x}W_xO_3$ ($0.08 \leq x \leq 0.4$) is presented.

## Introduction:

A plethora of research activities reported over the last several decades have established the significance of the charge density wave (CDW) and spin density wave (SDW) ground states for the correlated-electron systems in transition metal oxide compounds. The study of materials exhibiting these *“density-waves”* is interesting due to its proximity with physical phenomena of basic and technological interest, like high-$T_c$ superconductivity[1-8], colossal magnetoresistance [9-15] and magneto-electric multiferroics [16-18], which are yet way apart from complete understanding. The CDW and SDW are periodic modulations of the charge-density and the magnetization of the spin-moments [19,20,21], being manifested as signatures of strong electron-lattice and electron-electron correlations of the valence electrons, respectively. The presence of CDW and charge-stripes in under-doped cuprates has been confirmed through various ways [1-7]. It resolves the role of el ectron-phonon coupling inducing CDW vis-a-vis superconductivity [22]. Iron pinictides are relatively recent examples which show coexistence of superconductivity and SDW [23-26], while in manganites the occurrence of charge-ordering is now well established as CDW [12,13,27,28,29]. Its sliding results magnetoresistance due to field-induced collective motion of the charge-order [12]. Also, the incommensurately modulated CDW and magnetic structures may break the inversion symmetry which results in improper–ferroelectrics [30]. The manganites like $RMnO_3$ (R= Gd, Tb, Dy, Y and Ho) [31-34] and $R_2Mn_2O_5$ (R= Tb, Dy and Ho) [35-37] show incommensurate helical magnetic structure caused by competing exchange interactions. In the present scenario, the CDW and SDW are important and desired instabilities to be studied in a variety of correlated transition-metal oxides.

The different spin and charge ordered phases in rare-earth (R) and alkaline-earth (M) based perovskite $(R_{1-x}M_x)MnO_3$ arise due to multi-valency of the Mn ions achieved through hole-doping by substitution of divalent alkaline-earths ($M^{+2}$) at A-site [27,38,39,40,41]. Further these systems undergo structural distortion, as a result of Jahn-Teller effect due to the presence of $e_g$-orbitals of $Mn^{+3}$, indicative of orbital-ordering (OO) and often CDW type commensurate and or incommensurate (IC) charge-ordering [42-45]. In these systems the charge and orbital order are usually followed by variety of anti-ferromagnetic (AFM) ordering. Despite so many reports on the occurrence of variety of AFM structures, there is no report of the occurrence of IC-SDW order in these manganites.

The multi-valency however can be achieved by electron doping at $Mn^{+4}$ site by some suitable substitute at the B-site as well [46,47]. But there are sparse reports [48,49] in this regard.

In the present study we have investigated the solid-solution series of $SrMn_{1-x}W_xO_3$ for $x$ = 0.08 to 0.1875. Here the $Mn^{+4}/Mn^{+3}$ ratios have been tuned by substitution of $W^{+6}$ ions at the Mn-site. Compared to the hole-doping route of tuning the $Mn^{+3}/Mn^{+4}$ ratio, the electron-doping route bears some differences. $W^{+6}$ being diamagnetic cuts the existing nearest neighbor (NN) Mn-O-Mn super-exchange pathways, which may affect the generic magnetic ground states in manganites. This is the first attempt to study the generic change in the magnetic ordering in electron doped manganites. Our detailed neutron powder diffraction (NPD) studies on this series show the occurrence of IC-CDW and IC-SDW phases, which get continuously tuned with W substitution.

**Experimental:**

$SrMn_{1-x}W_xO_3$ (x = 0.08 to 0.1875) were prepared through the conventional solid state reaction route using 99.99% pure $SrCO_3$, $MnO_2$ and $WO_3$. The stoichiometric mixture of the ingredients was thoroughly ground mixed using mortar-pestle for ~ 5 h and then calcined at 1200 °C for 24 h. The calcined powder was reground and pelletized in the form of ∼ 14 mm × 1 mm disks and then sintered at 1450 °C for 12 h. The x-ray diffraction (XRD) measurements have been done using Cu-Kα radiation on a Rigaku diffractometer (D-max) equipped with a graphite (002) monochromator. Electron diffraction studies have been performed using transmission electron microscopy equipped Gatan high-temperature (model 652) and liquid-nitrogen based low-temperature holders (model 636MA). The dc-magnetization measurements as a function of temperature (*M-T*) and field (*M-H*) measurements were carried out between 2 and 400 K, using SQUID-VSM (QD). The neutron powder diffraction (NPD) has been performed in time-of-flight (TOF) mode at GEM beamline of ISIS facility, UK. Structural refinements of the XRD and NPD data has been performed using JANA 2006 [50].

**Results:**

**Room temperature structural studies**

Room-temperature (RT) structural analysis and phase purity confirmation of the as-prepared samples was done through TEM, XRD and TOF-NPD measurements, followed by structural refinement of the XRD and NPD data. For details, see section I of the supplementary materials. At room temperature all the $SrMn_{1-x}W_xO_3$ samples are found to be single phase having tetragonally distorted perovskite structure with space-group *P*4/*mmm*, while the tetragonality ($c/a$) changes from elongated ($c/a > 1$), for $0.08 \leq x < 0.1$, to compressed ($c/a < 1$), for $0.1 \leq x < 0.1625$. The compositions with $0.1 \leq x < 0.1625$ were found to undergo incommensurate charge-order modulation (IC CO) along the direction close to [110]. Figure 1(a) presents a representative example of a <001> zone SAED pattern showing RT IC modulation along ~ [110] in $SrMn_{1-x}W_xO_3$ samples, e.g. $x$ = 0.15. Figure 1(b) illustrates splitting behavior of (200)/(020) and (002) XRD peaks and thus demonstrates the variation and the crossover of the tetragonality as a function of '$x$'. The $x$ = 0.1 composition appears to be the critical boundary, across which the tetragonality crossover occurs and therefore it exhibited coexistence of both type of tetragonal phases (elongated + compressed). Two-phase Rietveld refinement of the XRD data for $x$ = 0.10 sample reveals 25 % of elongated ($c > a$) and 75 % of compressed ($c < a$) tetragonal phases at RT, see Fig. SM1(b) of the supplementary materials. However, with decreasing temperature the relative fraction of the elongated-tetragonal phase increases at the expense of IC CO modulated compressed-tetragonal phase and below ~ 200 K only elongated-tetragonal phase persists. The sample with x = 0.175 crystalizes in cubic perovskite at/above room temperature. However, it transforms to IC CO superlattice modulated compressed-tetragonal phase below 250 K, see Fig. SM6(a-c) of supplementary materials.

Figure 1(c) shows the RT-XRD profiles with IC CO superlattice peaks occurring symmetrically about the (200/020) basic perovskite peaks. The dashed lines are a guide to the eyes, showing the variation in the modulation vector as a function of '$x$'. The wavelength of the modulation (**m**) of the IC ordering along ~ [110] is estimated directly from the measured XRD profiles [51] and is given by equation (1).

$$m = 2|\mathbf{g}|^2 \Big/ \left( |\mathbf{p}|^2 - |\mathbf{q}|^2 \right) \tag{1}$$

where, **g** is the reciprocal vector corresponding to the basic perovskite peak (200) and **p** and **q** are the reciprocal vectors corresponding to the superlattice peaks symmetrically appearing on both sides of the (200) peak. The rough estimate of the modulation (**m**) was further refined following the total profile fitting including the CO superlattice peaks, using higher-dimensional ((3+1)d) Le-Bail fitting, see Figure SM4 of supplementary materials. The positions of the incommensurately modulated peaks for the sample with $x$ = 0.1375 matched best with the superspace group $P2/m(\alpha\beta0)00$ with CO modulation vector $\mathbf{q}_{IC} = (\alpha a^*+\beta b^*+0)$ to be (0.112, 0.122, 0). In fact, on CO modulation the *P4/mmm* basic phase undergoes a nominal monoclinic distortion with superspace group $P2/m(\alpha\beta0)00$ and therefore $\alpha$ and $\beta$ slightly differ, i.e. the modulation direction is not exactly [110]. The value of 'm' was calculated as $m = g_{110}/|\mathbf{q}_{IC}|$. The refined values of the lattice parameters, the modulation vectors and the related space-groups and superspace-groups are listed

in Table 1. Figure 1(d) shows that 'm' decreases with increasing W content, i.e. with increasing $Mn^{+3}/Mn^{+4}$ ratios.

**dc-Magnetization studies**

Figure 2 (a-d) shows temperature dependence of *dc*-magnetic susceptibility ($\chi_{dc}$-*vs*-*T*) of $SrMn_{1-x}W_xO_3$ for x = 0.08, 0.10, 0.1375 and 0.1625 respectively. Each composition exhibits two anomalies, one above RT ($T_1$) and the other below RT ($T_2$). The anomaly at $T_1$ appears to be originated from magneto-elastic coupling. For x = 0.08, $T_1$ is cubic (*Pm*-3*m*) to tetragonal (*P*4/*mmm*) transition temperature whereas for x = 0.10, 0.1375 and 0.1625, it is associated with CO transition ($T_{CO}$). The second anomaly at $T_2$ can be identified as ($T_N$), corresponds to paramagnetic (PM) to AFM transition. It is observed that while $T_N$ decreases but the $T_{CO}$ increases with increasing '*x*'. Isothermal field dependent magnetization (*M-H*) measurements were also carried out at various temperatures, see Figure.SM8 of the supplementary material. At higher temperatures the *M-H* shows typical PM behavior (linear). Below $T_N$, the *M-H* still remains linear but with much lower magnetization values and therefore it is attributed to AFM phase. The occurrence of AFM orders were finally confirmed through NPD measurements as described in the following.

**Neutron diffraction studies**

Figure 3(a) shows the room temperature TOF-NPD profiles (bank-4) of $SrMn_{1-x}W_xO_3$ for compositions with x = 0.08 to 0.1875. Occurrence of CO superlattice peaks is more prominent in NPD as indicated by the bent arrows. The composition with x = 0.08 does not show any superlattice peaks in NPD, consistent with XRD. The CO superlattice peaks start appearing for x ≥ 0.1 and persists upto '*x*' = 0.1875, as marked by the dotted line. Unlike XRD, weak and diffuse CO superlattice peaks can be seen for *x* = 0.1875. High temperature NPD was also performed for sample with *x*=0.08 and 0.1375, which shows cubic *Pm*-3*m* structure, see Figure. SM10 of supplementary materials. Figure 3(b) presents low-temperature (5-10 K) bank-3 TOF-NPD profiles of $SrMn_{1-x}W_xO_3$ for *x* = 0.08 to 0.3. The region between 4.7 to 6.5 Å, as circumscribed under the dotted rectangle, highlights the main magnetic peaks appearing due to AFM order. It can be clearly noticed that drastic changes occur in the magnetic ground state of $SrMn_{1-x}W_xO_3$ from x=0.08 to 0.3. It exhibits four different magnetic phases, with boundaries separated by only slight variation in the W content. For these four magnetic phases, we could identify the composition boundaries as 0.08 ≤ *x* ≤ 0.10, 0.125 ≤ *x* ≤ 0.15, 0.1625 ≤ *x* ≤ 0.175 and 0.2 ≤ *x* ≤ 0.40. The fourth composition-range basically exhibits spin-glass phase, whose details is discussed elsewhere [52]. To understand the differences between the spin structures of hole doped and electron doped AFM manganites, the magnetic structures of the observed AFM phases in $SrMn_{1-x}W_xO_3$ have been analyzed using Rietveld refinement of the NPD data.

Table 2 lists the nuclear and magnetic unit cell parameters along with the corresponding propagation vectors **k**. For $x$ = 0.08 and 0.10, the **k** is (1/2, 1/2, 0) w.r.t. to the elongated-tetragonal basic perovskite lattice. Magnetic structure was solved through Rietveld refinement using Jana-2006, see Fig. 4(a) and 4(b). The inset of Fig. 4(b) depicts the unit cell obtained from the refinement. This shows a ferromagnetic (FM) interaction along *c*-axis, i.e. along the 4-fold axis of the tetragonal phase, while the nearest neighbor (NN) interaction between the chains, in the *ab*-plane, is anti-ferromagnetic. This is a typical *C*-type AFM structure [53,54]. The refined values of the Mn moments for x = 0.08 and 0.10 are found to be 2.27 and 2.25$\mu_b$, respectively, which is quite in agreement with other reports for $Mn^{+4}$ spin only moments obtained through NPD studies [55,56,57]

Interestingly, the AFM order for $x$ = 0.1375 and 0.15 could be characterized by minimum of two propagation vectors, one commensurate, $\mathbf{k}_C$, and the other incommensurate, $\mathbf{k}_{IC}$. For $x$ = 0.1375, the value of $\mathbf{k}_C$ and $\mathbf{k}_{IC}$ are = (1/2, 0, 1/2) and (0.12, 0.38, 1/2) whereas for x= 0.15, the values are (1/2, 0, 1/2) and (0.13, 0.37, 1/2) respectively, w.r.t. to the compressed tetragonal basic perovskite lattice. The vector $\mathbf{k}_C$ = (1/2, 0, 1/2) was able to account for most of the observed magnetic peaks, except for the peaks at d = 4.8 and 6.08 Å. These peaks could be accounted only by the IC second propagation vector $\mathbf{k}_{IC}$. The FM chain for this new "*C*-type AFM" is along 2-fold axis in contrast to the 4-fold axis of the *P*4/*mmm* basic perovskite for x = 0.08 and 0.10. The AFM order for $SrMn_{1-x}W_xO_3$ with x = 0.1625, could be characterized by single IC propagation vector, $\mathbf{k}_{IC}$ = (0.07, 0.43, 1/2).

Keeping in view, the existence of two propagation vectors $\mathbf{k}_C$ and $\mathbf{k}_{IC}$ for x= 0.1375 and 0.15, two possibilities exists. Firstly, that the net magnetic phase is a composite of two magnetic phases coexisting below $T_N$, each characterized with different vectors. Another possibility is that the whole bulk consists single magnetic phase in which the spin-structure is characterized by two propagation vectors. Since the magnetic peaks corresponding to both propagation vectors arise simultaneously, therefore the second possibility seems to be more appropriate to explain the observed data. For the compositions with x = 0.1375 (and similarly for x=0.15), 18 possible irreps solutions were suggested by ISODISTORT [58], out of which only 3 could fit the observed data. The goodness of fit (GOF) parameter for these three structures is almost identical (~2.68). The NPD data with the best possible nuclear and magnetic refinement is shown in Fig. 5 and the corresponding refined magnetic structures are shown in Fig. 6(a-c). From Fig.6 (a-c) it is quite clear that the value of magnetic moments at each Mn-site are not the same, these are changing periodically giving rise to a SDW structure. The exhaustive list of irreps generated for the phases with x = 0.1375 and 0.1625 is given in Table 3.

The $x$ = 0.1625 sample is characterized by IC vector k= (0.07, 0.43, 1/2), two possible model seems appropriate to fit the NPD data with almost similar value of GOF (~2.2). Out of these two solutions, one is centrosymmetric {P2/m.1'(ab1/2)00s} and the other is non-centrosymmetric {Pm.1'(ab1/2)0s}. Figure 7 (a) shows the Rietveld refined NPD data for x = 0.1625 composition. Figures 7(b-c) shows respectively the centrosymmetric and non-centrosymmetric magnetic structure models resulted from the refinement. The centrosymmetric model exhibits a collinear sinusoidal SDW (Fig. 7b) whereas the non-centrosymmetric model (Fig. 7b) shows a cycloidal SDW confined in the *ab*-plane. It is difficult to conclude the correct model just on the basis of GOF. Since the magnetic structure is fitted very well with the non-centrosymmetric space-group, therefore it is likely to be a promising candidate for multiferroic studies like $TbMnO_3$, DyMnO3 and $EuMnO_3$.

**Discussion:**

**Charge density wave (CDW) modulation**

The $SrMnO_{3-\delta}$ forms basic perovskite with Mn existing in +4 charge-state, whose moments have *G*-type AFM order [55]. Substitution of $W^{+6}$ at $Mn^{+4}$ site effectively acts as electron doping to $Mn^{+4}$ and a proportionate fraction of $Mn^{+4}$ gets convert to $Mn^{+3}$, creating mixed valance state. Such mixed Mn charge states have been reported for $SrMn_{1-x}Mo_xO_3$ [59,60]. In perovskite manganites, charge-order superlattice modulation occurs along [110] of the basic perovskite. The modulation occurs due to ordering of $e_g$-electrons of $Mn^{+3}$ in a $Mn^{+3}/Mn^{+4}$ double-valent state of Mn [39]. When $Mn^{+3}/Mn^{+4}$ ratio is rational (commensurate) like 1/1, 1/2 and 1/3, the modulation is commensurate and well-defined charge-strips [61], optimized through long ranged coulomb interaction, giving rise to Wigner like crystallization [62]. The modulation manifests itself as a periodic deformation of the $MnO_6$ octahedra, arising due to the Jahn-Teller distortion (JTD) caused by $Mn^{+3}$ ions. In the case of $Fe_3O_4$ it has been experimentally [63] and theoretically [64] shown that CO may result by only ~20% charge-disproportionation i.e. the difference in the ordered charge states may be only 0.2e and not 1e. The ordering of the disproportionate charge results as a CDW instability in the presence of electron-phonon interaction [65]. In the present case also the CO can be described as $Mn^{+4-\psi}$ /

$Mn^{+4+\psi}$, where the charge-disproportionation $\psi$ gets continuously and periodically distributed over the constant background of $Mn^{+4}$ lattice, such that the period is described by the modulation vector $\mathbf{q}_{IC}$. The charge-disproportionation $\psi$ arises here due to appearance of $e_g$-electrons through $W^{+6}$ substitution. For $SrMn_{1-x}W_xO_3$ with $x$ = 0.15 the average charge state of Mn will be ~ 3.7 hence the charge-disproportionation will vary as $0 \leq \psi \leq 0.3e$. The conceptualization of charge-disproportionation remains rather hidden in the case of commensurate CO. But the occurrence of incommensuration and the temperature dependence of the modulation vector, as observed in the present case, make it vivid. Since the extra charge $\psi$ has $e_g$-character, its disproportionate distribution at Mn sites will cause equally disproportionate JTD, i.e. the deformation of $MnO_6$ octahedra and the corresponding shift in the oxygen positions will exactly follow the charge-disproportionation $\psi$. In the case of IC modulation the deformation and shift in oxygen positions do not remain pinned with the Mn sites, rather change continuously and independently along the vertex connected $MnO_6$ octahedra-chain. This, in turn, translates that the $e_g$-charge, which causes lattice deformation, has some itinerant character [66] as well, and therefore, it spreads over the vertex connected $MnO_6$ octahedra square net-work as continuous charge density wave (CDW). CDW picture is the most general representation of CO modulation. The CDW like character of the CO phase has already been discussed earlier [12] and is also reported [27,28] recently for hole-doped manganites. In the present report we show its presence in electron doped manganites. The occurrence of planer ordering $3\boldsymbol{d}_{x^2-y^2}$ orbitals play vital role in the CDW instability in the present case.

## Planer ordering of $3d_{x^2-y^2}$ orbitals

The JT-active $Mn^{+3}$ ion causes structural and magnetic anomalies in $SrMn_{1-x}W_xO_3$. The $e_g$-charge of the $Mn^{+3}$ ion can either occupy $3\boldsymbol{d}_{z^2}$ or $3\boldsymbol{d}_{x^2-y^2}$ or a mixed state as represented by $\langle\psi\rangle = \alpha\langle \boldsymbol{d}_{z^2}\rangle + \beta\langle \boldsymbol{d}_{x^2-y^2}\rangle$ [67]. At high temperatures, in the cubic perovskite phase, the $e_g$-charge keeps hopping and both the orbitals are equally probable. As temperature decreases it starts localizing and then only one of the orbitals dominates, leading to structural distortion due to JT-effect. Figure 8(a) shows splitting of the (200) peak of the cubic perovskite of $SrMn_{1-x}W_xO_3$ with x = 0.08. The behavior of splitting indicates that at lower temperatures it goes to an elongated tetragonal distortion, i.e. $c > a$. It means that the $e_g$-charge totally occupies the $3\boldsymbol{d}_{z^2}$ orbitals, resulting in its ferro-ordering, as shown in Figure 8(e). Such an orbital-order is known [39] to result *C*-type AFM order and the same has been observed in the present case as well. On the other hand, the $SrMn_{1-x}W_xO_3$ samples with $0.1375 \leq x \leq 0.175$, which undergo CDW modulations, invariably show compressed tetragonal distortion, i.e. $c < a$, see Figures 8(b, c and d). A compressed tetragonal distortion directly implies that the $e_g$-charge pre-dominantly occupies the $\mathbf{3d_{x^2-y^2}}$ orbitals having ferro-ordering in the *ab*-plane, as shown in Figure 8(f). Here it should be noted that the behavior of tetragonal splitting of (200/020) and (002) peaks during cooling are distinctly different for elongated and compressed tetragonal phases. For the elongated (x = 0.08) tetragonal phase the rate of temperature variation of the peak positions of (200/020) and (002) peaks is almost symmetric, whereas for the compressed (x= 0.1375, 0.1625 and 0.175) tetragonal phases, it is highly asymmetric. This has been graphically presented as temperature dependence of $d_{200}$ and $d_{002}$ in Fig.SM7 of supplementary materials.

The $SrMn_{1-x}W_xO_{3-\delta}$ composition with x = 0.1 is an intermediate composition with CDW ordering at RT. It has been observed that the tetragonal distortion continuously changes from elongated to compressed with '*x*'; see section-I of the supplementary materials. The CDW modulation vanishes as the compressed tetragonal distortion changes to elongated one at low enough temperatures. This directly implies that in the present case the occurrence of CDW ordering is directly related with the mode of compression of $MnO_6$ octahedra.

In conventional CO manganites the $MnO_6$ octahedra are usually elongated and tilted, resulting in structures with space groups like *I*4/*mcm*, *Pnma* or *Pbnm* [68]. Such space groups in perovskites give rise to distinctly new peaks in their XRD and NPD profiles at RT. But in the present case we did not observe any of those peaks, which could have indicated the possible existence of octahedra tilt ordering. All the perovskite peaks were accounted to the compressed tetragonal structure with *P*4/*mmm* symmetry and the remaining peaks were accounted to the incommensurate CDW structure with propagation vector $q_{ic} = (\alpha a^* + \beta b^* + 0)$ and superspace group P2/*m*($\alpha\beta0$)00, as listed in Table 1. This proves that the occurrence of planer ferro-ordering of $\mathbf{3d_{x^2-y^2}}$ orbitals in $SrMn_{1-x}W_xO_{3-\delta}$ with x ≥ 0.1375 has a well-defined physical correlation with the CDW. In $La_{0.5}Sr_{1.5}MnO_4$, the CO phase coexists with $\mathbf{3d_{x^2-y^2}}$ orbital-ordering [69] but in that case the orbital-ordering is not *ab*-planer type, rather the *xy*-plane of $\mathbf{3d_{x^2-y^2}}$ orbital is tilted normal to the crystal-lattice *ab*-plane, which contained the *CE*-type CO. The present observation of planer ferro-orbital-ordering of $\mathbf{3d_{x^2-y^2}}$ orbitals, co-existing with co-planer IC-CDW in $SrMn_{1-x}W_xO_{3-\delta}$ with x ≥ 0.1375, is unique and has not been observed so far in manganites.

Planer ferro-orbital ordering of $\mathbf{3d_{x^2-y^2}}$ orbitals is usually supposed to be metallic and, therefore, it should suppress charge localization and, hence charge-ordering should not appear. But there are well celebrated materials [70], like high-$T_c$ cuprates, where $3d_{x^2-y^2}$ orbitals, derived from $Cu^{2+}$ JT active ions, undergo planer-ordering on a square planer Cu-O lattice and gives rise to AFM insulating CDW phase. $SrRuO_2$ is another example, which shows $3d_{x^2-y^2}$ orbital-ordering with AFM insulating phase [71]. For stabilization of CDW Fermi-surface nesting should exist. The occurrence of planer orbital-order of $3d_{x^2-y^2}$ provide necessary precursor for the formation of $e_g$-electron band with nested Fermi-surface, resulting in CDW instability. The necessarily required lattice distortion arises due to electron-lattice interaction derived through JT effect. Here the 'itinerant' and 'localized' character of $e_g$-electron, manifest simultaneously [66].

**Spin density wave (SDW) ordering**

$SrMn_{1-x}W_xO_{3-\delta}$ with $x \leq 0.10$ resulted in *C*-type AFM phase with propagation vector (1/2, 1/2, 0) under ferro-ordering of $3d_{z^2}$ orbitals, but the $SrMn_{1-x}W_xO_{3-\delta}$ compositions with x = 0.1375 and 0.15 resulted in three possible magnetic structures, which are defined by the superposition of the commensurate (1/2, 0, 1/2) and incommensurate (0.12, 0.38, 1/2) vectors under planer ordering of $3d_{x^2-y^2}$ orbitals with CDW. A close inspection of the models in Figure 6(a, b) indicates that the canting angles between NN moments are ranging from 30° to 110°. Following the Dzyaloshinskii-Moriya (DM) interaction moment configurations with such large canting angles demand high Mn-O-Mn bond angles. But the crystallographic analysis for these phases has already shown that the basic perovskite structure of these compounds is *P*4/*mmm*, which bears Mn-O-Mn bond angles to be 180° and hence it cannot allow such large canting angles. Of course the CDW modulations do lead to some minor bending of the Mn-O-Mn bonds, but it is only few degrees and hence insufficient to cause large canting between NN Mn moments. Rietveld refinement of the profile in Fig.9 (a) shows that the propagation vector $k_{IC}$ = (1/2, 0, 1/2) fits all the magnetic peaks except for the two peaks at d ~ 4.8 and 6.08 Å. The corresponding magnetic model is shown in Figure 9(b). A comparison between the models in Figure 9(b) and Figure 6(c) clearly shows that they have nearly identical orientation of the moments. The effect of superimposition of the second IC propagation vector is simply to modulate the amplitudes of the co-linearly aligned AFM moments, and to produce the SDW structure. Thus, the SDW model shown in Figure 6(c) is the correct model for the $SrMn_{1-x}W_xO_{3-\delta}$ with x = 0.1375 and 0.15. The refined values of magnetizations are shown in Table SM1 of the supplementary materials. These parameters show that the Mn-ion magnetization under goes sinusoidal modulation of the co-linear moments in the SDW and at each site the Mn moments have two parts. The fixed part constantly prevails at each Mn site and the other part changes sinusoidaly, thus modulating the magnetization at each site. The sinusoidal spatial variation of the net magnetization and its $\mathbf{M_x}$ and $\mathbf{M_y}$ components are shown in Fig. SM12 of the supplementary materials.

Out of the two possible SDW structure solutions for the compositions with $0.1625 \leq x \leq 0.175$ the one with parallel moments, as shown in Figure 7(b), appears to be the most likely. However, the small bending of Mn-O-Mn bonds, appearing due to CDW modulation, can still account for the observed small ($10^o$-$15^o$) canting angles between NN moments, as shown in Figure 7(c), and hence this model is also quite likely. The final decision for $x = 0.1625$ sample can only come from proper energy calculation for these two models, which is currently beyond the scope of this work. Table SM1 of the supplementary materials shows the refined values of the magnetization for these models. Unlike that of x = 0.1375 and 0.15 samples, this model does not have any fixed-moment component. Here, whole of the Mn-ion magnetization under goes sinusoidal modulation, resulting in co-linear and non-collinear AFM moments in the SDW.  The sinusoidal spatial variation of the net magnetization and its $\mathbf{M_x}$ and $\mathbf{M_y}$ components are shown in Fig. SM13 of the supplementary materials.

Keeping in view that the magnetic phases at the two extremes are being characterized by just single propagation vectors, the one with $x \leq 0.10$ by the commensurate vector (1/2, 1/2, 0) and the one with $0.1625 \leq x \leq 0.175$ by the incommensurate vector (0.07, 0.43, 1/2), the magnetic phases of the intermediate compositions with $0.125 \leq x \leq 0.15$, which are characterized by two vectors, arise due to competing magnetic interactions.

**Frustration induced SDW order**

A SDW modulation arises due to position dependent magnetization of moments [21]. The CDW and SDW are usually observed in systems where Fermi surface nesting exists. The incommensurate SDW was first observed in metallic chromium due to Fermi surface nesting [20]. However, the existence of Fermi surface nesting is not a necessary and sufficient criterion for SDW. There are examples of insulating oxide systems, which show SDW originating due to magnetic frustration [31-37]. Spin-orbit coupling plays a crucial role in deciding moment frustration [72].

Since in diffraction experiment the sharp peaks appear due to elastically scattered neutrons from the average common coherence of the structural order; just from elastic peaks one cannot differentiate between the time-averaged (due to moments still under partial order/paramagnetic state) and space-averaged (due to moments undergone partial ordering and still accompanying frozen disorder) configurations of moments, that which one is actually participating in the spatial variation of the magnetization in a SDW structure. One will have to take help of diffuse part of the elastic scattering. As discussed in the following, the present observation appears to be a case of moment configuration resulting from gradual mixing of frustration with order.

Figure 10 shows stack of NPD profiles highlighting the region having major magnetic peaks. It can be noticed that as W increases a broad humped region slowly grows over the flat background. For $x \leq 0.15$ the underlying hump is negligible. But with increasing W i.e. for $x \geq 0.1625$, the broad hump centered ~ 5.6 Å, significantly increases and maximize for $x = 0.1825$. With further increase of W i.e. for $x > 0.1825$, the broad hump starts decreasing and almost vanishes for $x = 0.30$. It was reported that the compositions with $x > 0.25$ belong to spin-glass phase [52] The ZFC and FC bifurcations supports the presence of frustration in the system. The ZFC and FC bifurcation is almost absent till $x = 0.1$ but it continuously increases for $x \geq 0.1375$. For $x = 0.1375$ it was nominal but for x= 0.1625 it becomes quite prominent, see Figure. SM9 of the supplementary materials. It indicates that out of the two suggested possible origins of the observed SDW ordering, the partially frustrated spin-order is physically more justified for the present case. The net ordered moment part gives sharp magnetic peaks, whereas the frustrated moment configuration, which has some degree of short range correlation, gives the broad hump centered ~ 5.6 Å in the NPD. When fraction of frustration maximizes the short-range AFM phase fully dominates the long-range SDW order at $x \geq 0.1825$. At extreme frustration, e.g. $x = 0.30$, even the short-range order disappears making the background totally flat. This indicates that the observed SDW modulation is arising due to frustration, which increases with '*x*', and finally leads to spin-glass phase.

Based on the above described magnetization and NPD studies, detailed magnetic and structural phase diagram of $SrMn_{1-x}W_xO_{3-\delta}$ has been constructed and presented in Fig. 11 as a function of W. The pristine $SrMnO_3$ is a *G*-type antiferromagnet [55**Error! Bookmark not defined.**] with cubic perovskite structure. On W doping i.e. with increasing '*x*', the Mn-O-Mn superexchange interaction changes, which results novel magnetic ground states. For $0.08 \leq x \leq 0.10$, the AFM structure changes from cubic *G*-type to an elongated tetragonal (c>a) *C*-type AFM with (1/2, 1/2, 0) propagation vector under ferro-ordering of $\mathbf{3d_z{}^2}$ orbitals. For $x \geq 0.1$, the magnetic interaction further changes to a different *C*-type AFM structure with k= (1/2, 0, 1/2), coexisting with a superimposed incommensurate magnetic propagation vector (0.12, 0.38, 1/2) resulting in a SDW. This SDW has a planer ferro-ordering of $\mathbf{3d_{x^2-y^2}}$ orbitals with CDW modulation of a compressed tetragonal (c<a) phase. For $0.1625 \leq x \leq 0.175$, the magnetic interaction gets redefined and now a single incommensurate propagation vector k= (0.07, 0.43, 1/2) prevails the SDW with CDW in a compressed tetragonal phase. For $0.175 \leq x \leq 0.2$, the system becomes disordered cubic perovskite which undergoes short-range AFM ordered. For compositions with $x > 0.2$, the system slowly transforms to a PODP phase, which undergoes spin-glass freezing at lower temperatures. The half doped systems i.e. $x = 0.5$ is an ordered double perovskite with AFM phase.

**Conclusions:**

In summary, based on our above described structural and magnetic studies we conclude that creation of mixed valent states $Mn^{4+}$ and $Mn^{3+}$ through electron doping of $SrMnO_3$, by substitution of W at Mn site in $SrMn_{1-x}W_xO_{3-\delta}$, results a variety of commensurate and incommensurate structural and magnetic ground states comprising CDW and SDW. For $0.08 \leq x \leq 0.10$ the AFM structure changes from cubic G-type AFM to an elongated tetragonal (c>a) C-type AFM with (1/2, 1/2, 0) propagation vector under ferro-ordering of $\mathbf{3}\boldsymbol{d_{z}^{2}}$ orbitals. With W compositions, increased above x=0.1, the tetragonal phase under gores elongated (c>a) to compressed (c<a) transition due to the changeover of ferro-ordering of $\mathbf{3}\boldsymbol{d_{z}^{2}}$ orbitals to ferro-ordering of $\mathbf{3}\boldsymbol{d_{x^2-y^2}}$ orbitals. The shrinking of *c*-axis (c<a) changes the magnetic interaction leading to a different *C*-type AFM structure, which results from flipping of (1/2,1/2,0) magnetic vector to (1/2, 0, 1/2). The ferro-ordering of $\mathbf{3}\boldsymbol{d_{x^2-y^2}}$ orbitals provide extended wave character to the $e_g$-electrons, which act as necessary precursor for Fermi-surfaces nesting leading to CDW ordering (0.112, 0.122, 0) and SDW with propagation vector (0.12, 0.38, 1/2) superimposed over the commensurate structure with k= (1/2, 0, 1/2). The occurrence of SDW and the incommensurate nature of both these ordering arise due to magnetic frustration. With increasing W i.e. for $0.1625 \leq x \leq 0.175$, the magnetic frustration further increases, changing the effective magnetic ordering to a single incommensurate propagation vector k= (0.07, 0.43, 1/2). For $0.175 \leq x \leq 0.2$ the system becomes disordered cubic perovskite with short-range AFM order. With further increase in W content, the magnetic frustration increases so much that it leads spin-glass phase.

**Acknowledgements:** Authors gratefully acknowledge Dr. A. K. Sinha and Dr. V. Ganesan for their constant support and encouragement. Dr. A. Banerjee and Dr. R. J. Choudhary are acknowledged for magnetic measurements. Experiments at the ISIS Neutron Source were supported by beam time allocation from the Science and Technology Facilities Council (U.K.). VP acknowledges Project of the Czech Science Foundation No. 18-10504S for developing of the Jana software.

**Figure captions:**

**Figure 1.** (a) Typical example of a selected area electron diffraction (SAED) pattern, showing charge-order modulation superlattice spots in $SrMn_{1-x}W_xO_{3-\delta}$ for $0.10 \leq x \leq 0.1625$. (b) Stack of XRD profiles showing tetragonal (200)/(002) peaks. The splitting behavior of these peaks shows cubic to tetragonal phase transformation as a function of W content. The corresponding inset shows the *c*/*a* ratio representing the change-over from elongated ($c > a$) to compressed ($c < a$) transition in the basic tetragonal perovskite lattice. (c) Stack of XRD profiles showing variation in CO modulation period as a function of W content. Its graphical representation is shown in (d).

**Figure 2.** ZFC and FC M-T variations of $SrMn_{1-x}W_xO_{3-\delta}$ for x = 0.08, 0.10, 0.1375 and 0.1625. M-T curves corresponding to each composition show two features as indicated by $T_1$ and $T_2$. The $T_1$ corresponds to CO (or CDW) ordering and the $T_2$ to magnetic ordering.

**Figure 3.** TOF-NPD profiles (a) of bank-4 showing room temperature CO modulation in $SrMn_{1-x}W_xO_{3-\delta}$ for x = 0.08, 0.10, 0.1375, 0.15, 0.1625, 0.175 and 0.1875 and (b) of bank-3 showing low-temperature anti-ferromagnetic (AFM) ordering peaks, as highlighted by the dotted-dashed rectangle.

**Figure 4.** Nuclear and magnetic Rietveld refinement of 10 K TOF-NPD profiles, showing *C*-type AFM phase for $SrMn_{1-x}W_xO_{3-\delta}$ samples with x = (a) 0.08 and (b) 0.10. The inset of 4(b) shows nuclear and *C*-type AFM structure of these two compositions.

**Figure 5.** Nuclear and magnetic Rietveld refined 5 K TOF-NPD profile of $SrMn_{1-x}W_xO_{3-\delta}$ with x = 0.1375. C and IC labels indicate the corresponding commensurate and incommensurate peaks, respectively.

**Figure 6.** The three possible magnetic structures obtained from the magnetic refinement of the 5 K TOF-NPD data of $SrMn_{1-x}W_xO_{3-\delta}$ with x = 0.1375, showing spin density wave (SDW) nature.

**Figure 7.** (a) Nuclear and magnetic Rietveld refined 10 K TOF-NPD profile of $SrMn_{1-x}W_xO_{3-\delta}$ with x = 0.1625, using just one IC propagation vector. (b) and (c) show the SDW type magnetic structures for sample with x = 0.1625, obtained for the possible centrosymmetric and non-centrosymmetric space groups, respectively.

**Figure 8.** Difference in the behavior of tetragonal splitting of (200/020) and (002) XRD peaks during cooling, (a) for the elongated tetragonal phase (x = 0.08) and (b), (c) and (d) for the compressed tetragonal phases x = 0.1375, 0.1625 and 0.175, respectively. The orbital ordered states for the elongated (x = 0.08) and compressed (x = 0.1375, 0.1625 and 0.175) tetragonal phases are shown in (e) and (f), respectively.

**Figure 9.** (a) Nuclear and magnetic Rietveld refined 5 K TOF-NPD data of $SrMn_{1-x}W_xO_{3-\delta}$ with x = 0.1375 using the (1/2, 0, 1/2) commensurate magnetic propagation vector only. The corresponding inset shows the refined nuclear and magnetic unit cell. (b) Magnetic moments as arranged in the *bc*-plane.

**Figure 10.** Stack of bank-3 10 K TOF-NPD profiles, highlighting the composition dependence of the broad hump in the vicinity of the main AFM peaks.

**Figure 11.** Magneto-structural phase diagram, showing occurrence of various structural and magnetic phases as a function of composition and temperature in $SrMn_{1-x}W_xO_{3-\delta}$.

**Table 1:** Space group, super space group and refined lattice parameters of CDW phases

| $SrMn_{1-x}W_xO_3$ (x) | 0.08 | 0.10 | 0.1375 | 0.15 | 0.1625 | 0.1725/0.1875 |
|---|---|---|---|---|---|---|
| **Spacegroup** | *P*4/*mmm* | *P*4/*mmm* | - | - | - | *Pm*-3*m* |
| **Super-Spacegroup** | - | - | *P*2/*m*(αβ0)00 | *P*2/*m*(αβ0)00 | *P*2/*m*(αβ0)00 | |
| **CDW vector (q)** | | | 0.112,0.122,0 | 0.124,0.129,0 | 0.148,0.136,0 | |
| **Lattice parameters (*a*, *b*, *c*, α, β, γ)** | | | 3.8865, 3.8927, 3.8408, 90, 90, 90.03051 | 3.8860, 3.8862, 3.8380, 90, 90, 90.025 | | 3.889 |

**Table 2:** Table depicting nuclear and magnetic structure parameters used for magnetic refinement

| $SrMn_{1-x}W_xO_3$ | Nuclear cell parameters (Å) | Nuclear space group | Magnetic Propagation vector | Magnetic cell parameters (Å) | Magnetic Space group |
|---|---|---|---|---|---|
| **x = 0.08** | *a* = 3.8063<br>*c* = 3.8912 | *P*4/*mmm* | (1/2,1/2,0) | *a* = 5.383,<br>*c* = 3.891 | *P*4/*mbm* |
| **x = 0.10** | *a* = 3.8176<br>*c* = 3.9148 | *P*4/*mmm* | (1/2,1/2,0) | *a* = 5.399,<br>*c* = 3.915 | *P*4/*mbm* |
| **x = 0.1375** | *a* = 3.8899<br>*c* = 3.8325 | *P*4/*mmm* | (1/2,0,1/2)<br>and<br>(0.12, 0.38, 1/2) | *a* = 7.7799,<br>*b* = 8.6982,<br>*c* = 7.6649,<br>$\alpha = \beta = 90^O$,<br>$\gamma = 153.44^O$ | *B*2/*m*.1'[*c*]($\alpha\beta$0)000 |
| **x = 0.1625** | *a* = 3.8944<br>*c* = 3.8291 | *P*4/*mmm* | ( 0.07, 0.43, 1/2) | *a* = 3.8944,<br>*c* = 3.8291 | *P*2/*m*.1'($\alpha\beta$1/2)00s |
| | | | | *a* = 3.8944,<br>*c* = 3.8291 | *Pm*.1'($\alpha\beta$1/2)0*s* |

**Table 3:** Table listing the magnetic space groups and the irreps generated by ISODISTORT for nuclear structure and corresponding propagation vectors listed in Table 2. Those which are shown in bold show the best fit with the experimental NPD data.

| $SrMn_{1-x}W_xO_3$ (X) | Shubnikov Spacegroup | Representations |
|---|---|---|
| **0.08 and 0.10** | $P_C$4/mmm, **$P_C$4/mbm** | mM1**+, mM2+** |
| | $P_C$4/mbm, $P_C$4/mmm | mM3+, mM4+ |
| | $P_C$4/nbm, $P_C$4/nmm | mM1-, mM2- |
| | $P_C$4/nmm, $P_C$4/nbm | mM3-, mM4- |
| | $C_a$mme, $C_a$mmm | mM5+, mM5- |
| | $P_B$nma, $P_B$mma<br>$P_B$2/n, $P_B$2/m | mM5+, mM5-<br>mM5+, mM5- |
| **0.1375** | **B2/m1'$_a$(αβ0)000, B2/m1'$_a$(αβ0)000,**<br>**B2/m1'$_a$(αβ0)000**, B2/m1'$_a$(αβ0)000<br>B2/m1'$_a$(αβ0)000, B2/m1'$_a$(αβ0)000<br>B2/m1'$_a$(αβ0)0s0, B2/m1'$_a$(αβ0)0s0<br>B2/m1'$_a$(αβ0)0s0, B2/m1'$_a$(αβ0)0s0<br>B2/m1'$_a$(αβ0)0s0, B2/m1'$_a$(αβ0)0s0<br>B2/m1'$_a$(αβ0)0s0, B2/m1'$_a$(αβ0)0s0<br>B2/m1'$_a$(αβ0)0s0, B2/m1'$_a$(αβ0)0s0<br>B2/m1'$_a$(αβ0)000, B2/m1'$_a$(αβ0)000 | **mE1lmR1+**, **mE1lmR1+**<br>**mE1lmR1+**, mE1lmR2+<br>mE1lmR2+, mE1lmR2+<br>mE1lmR4+, mE1lmR4+<br>mE1lmR4+, mE2lmR1+<br>mE2lmR1+, mE2lmR1+<br>mE2lmR2+, mE2lmR2+<br>mE2lmR2+, mE2lmR4+<br>mE2lmR4+, mE2lmR4+ |
| **0.1625** | **P2/m1'(αβ1/2)00s**, P2/m1'(αβ1/2)00s<br>**Pm1'(αβ1/2)0s**, Pm1'(αβ1/2)0s | **mE1**, mE1<br>**mE2**, mE2 |

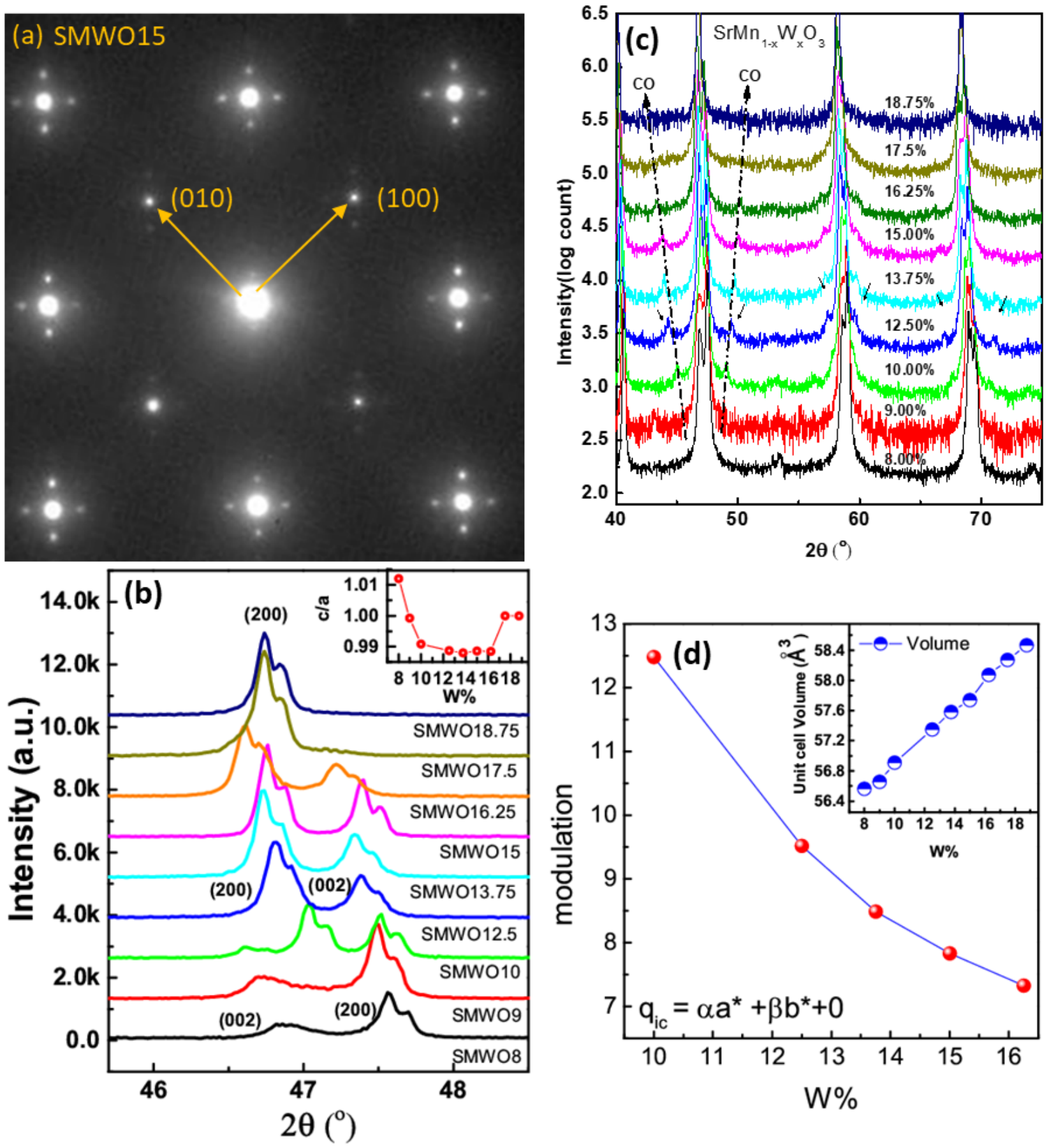
(a) SMWO15
(010)
(100)
(b)
Intensity (a.u.)
2θ (°)
SMWO18.75
SMWO17.5
SMWO16.25
SMWO15
SMWO13.75
SMWO12.5
SMWO10
SMWO9
SMWO8
(200)
(002)
c/a
W%
(c)
$SrMn_{1-x}W_xO_3$
CO
Intensity(log count)
2θ (°)
18.75%
17.5%
16.25%
15.00%
13.75%
12.50%
10.00%
9.00%
8.00%
(d)
modulation
W%
$q_{ic} = \alpha a^* + \beta b^* + 0$
Volume
Unit cell Volume (Å$^3$)

**Figure 1**

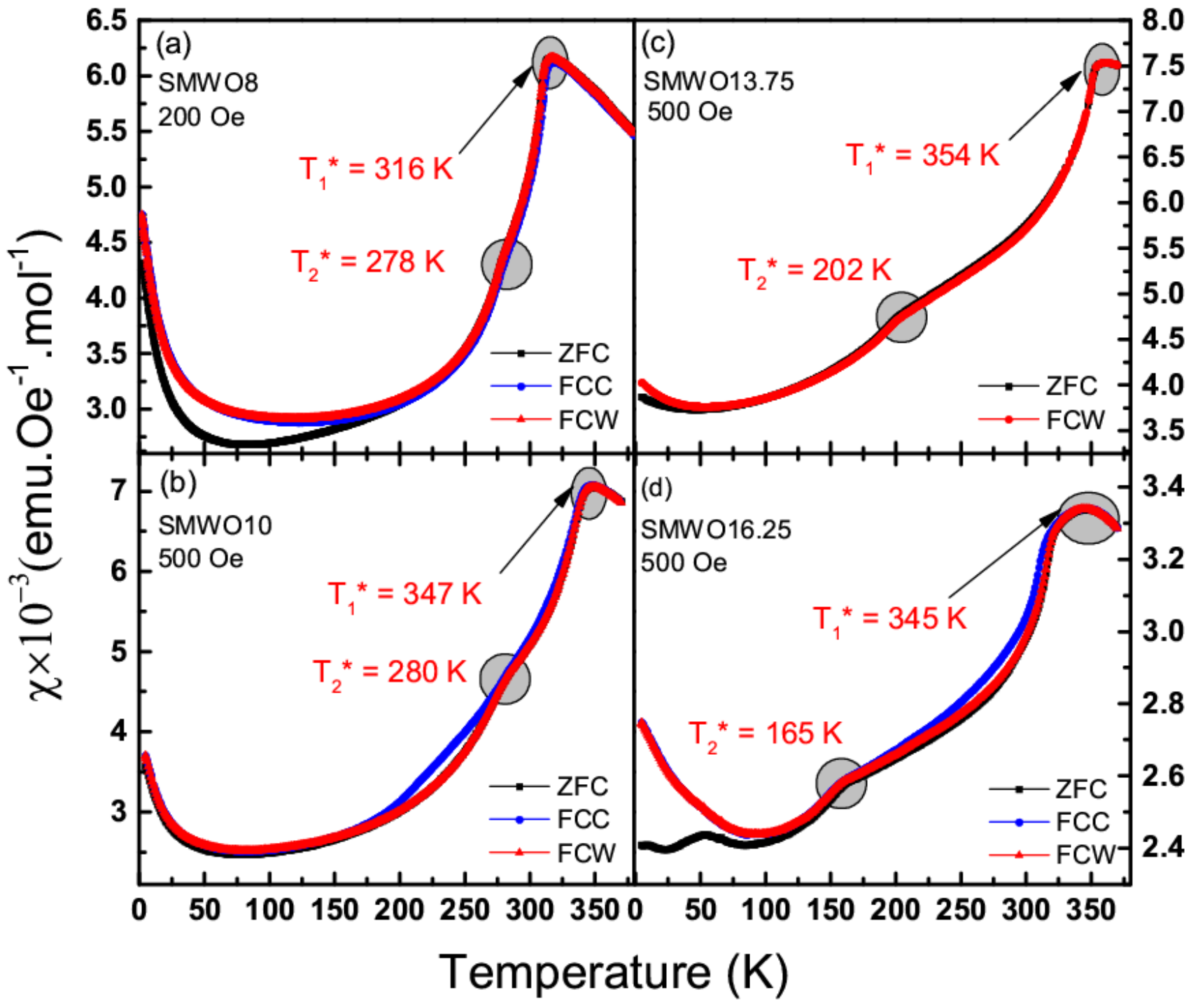
(a)
SMWO8
200 Oe
T1* = 316 K
T2* = 278 K
ZFC
FCC
FCW
(c)
SMWO13.75
500 Oe
T1* = 354 K
T2* = 202 K
ZFC
FCW
(b)
SMWO10
500 Oe
T1* = 347 K
T2* = 280 K
ZFC
FCC
FCW
(d)
SMWO16.25
500 Oe
T1* = 345 K
T2* = 165 K
ZFC
FCC
FCW
χ×10−3(emu.Oe−1.mol−1)
Temperature (K)

**Figure 2**

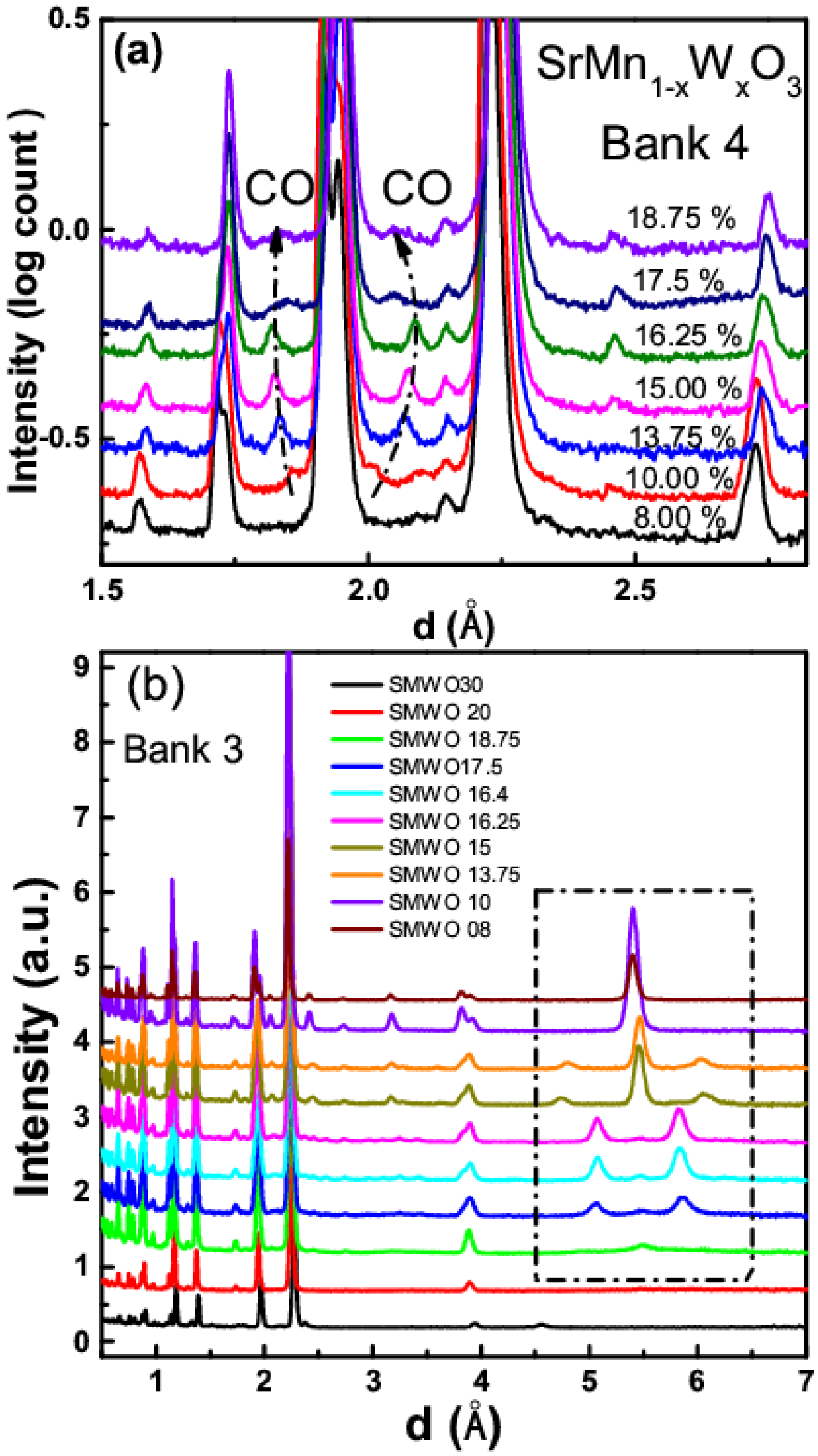

**Figure 3**

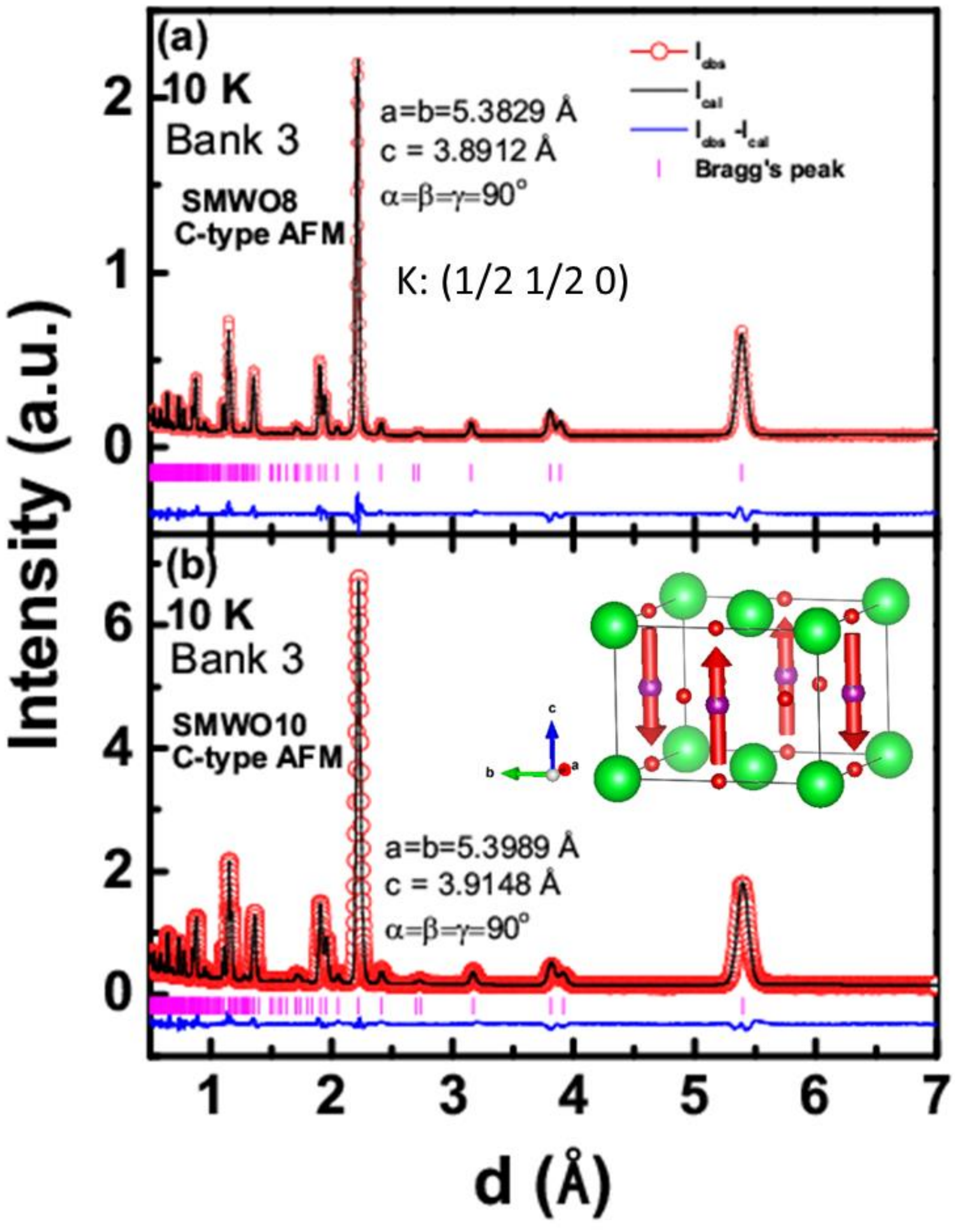

**Figure 4**

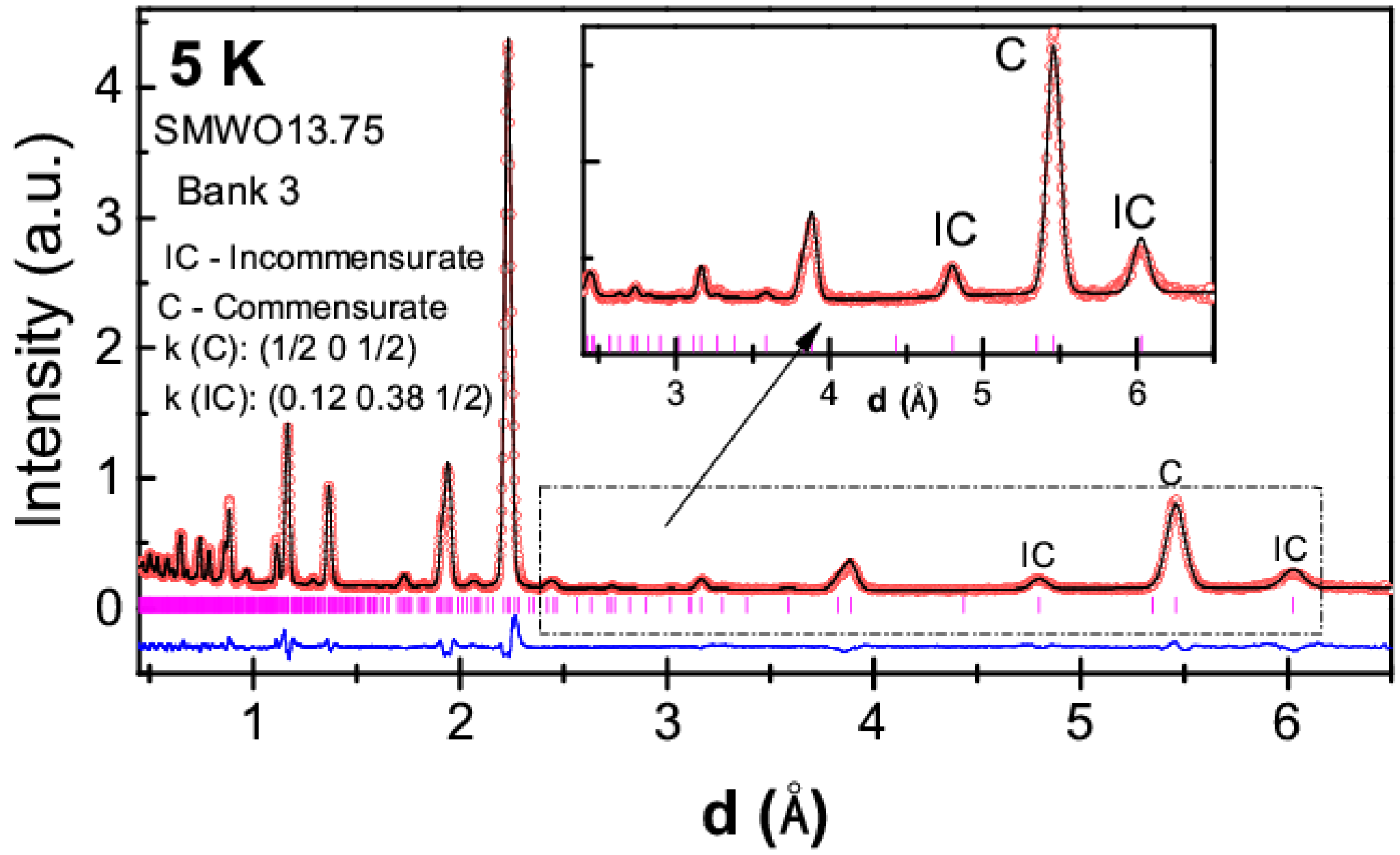

**Figure 5**

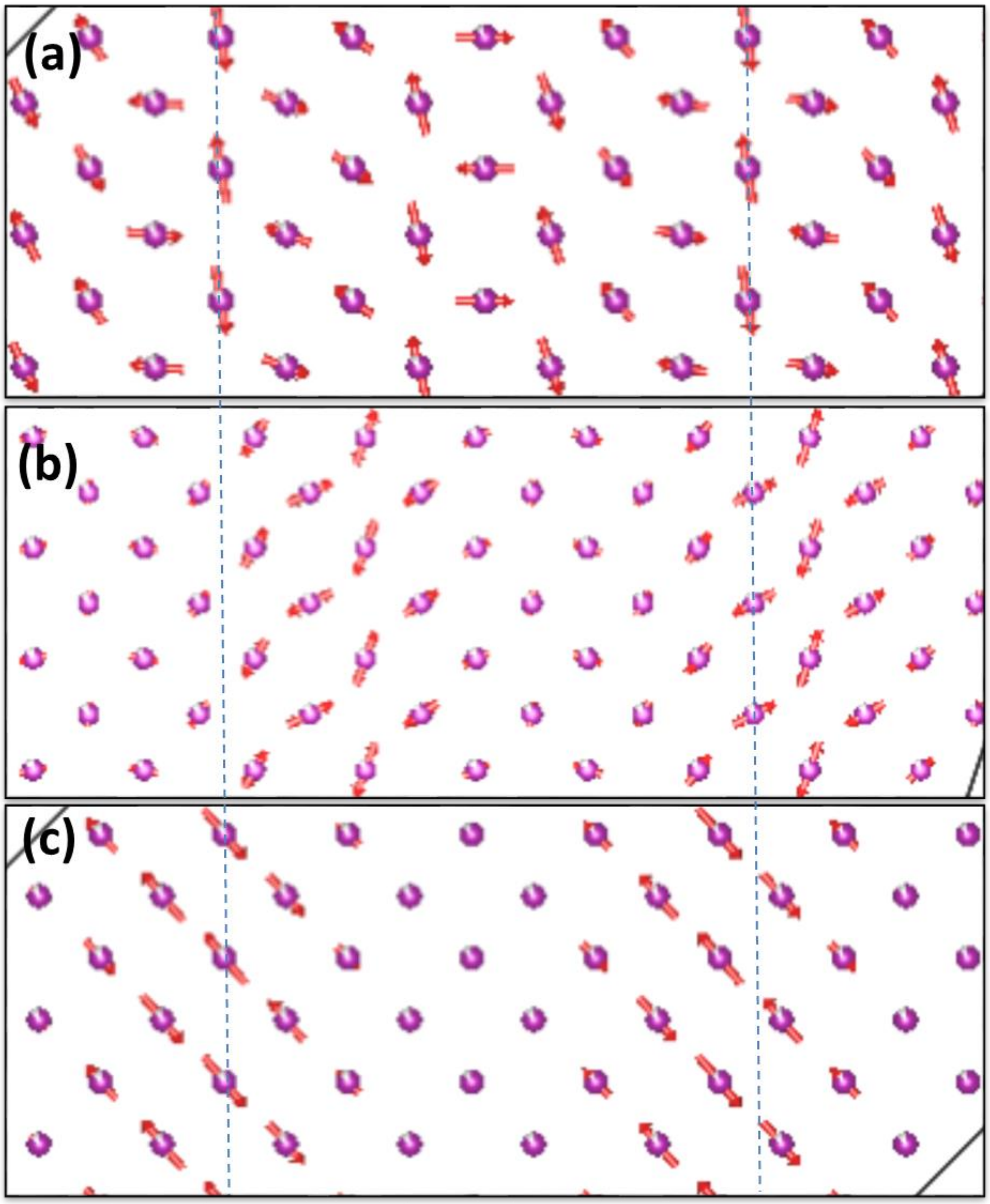

**Figure 6**

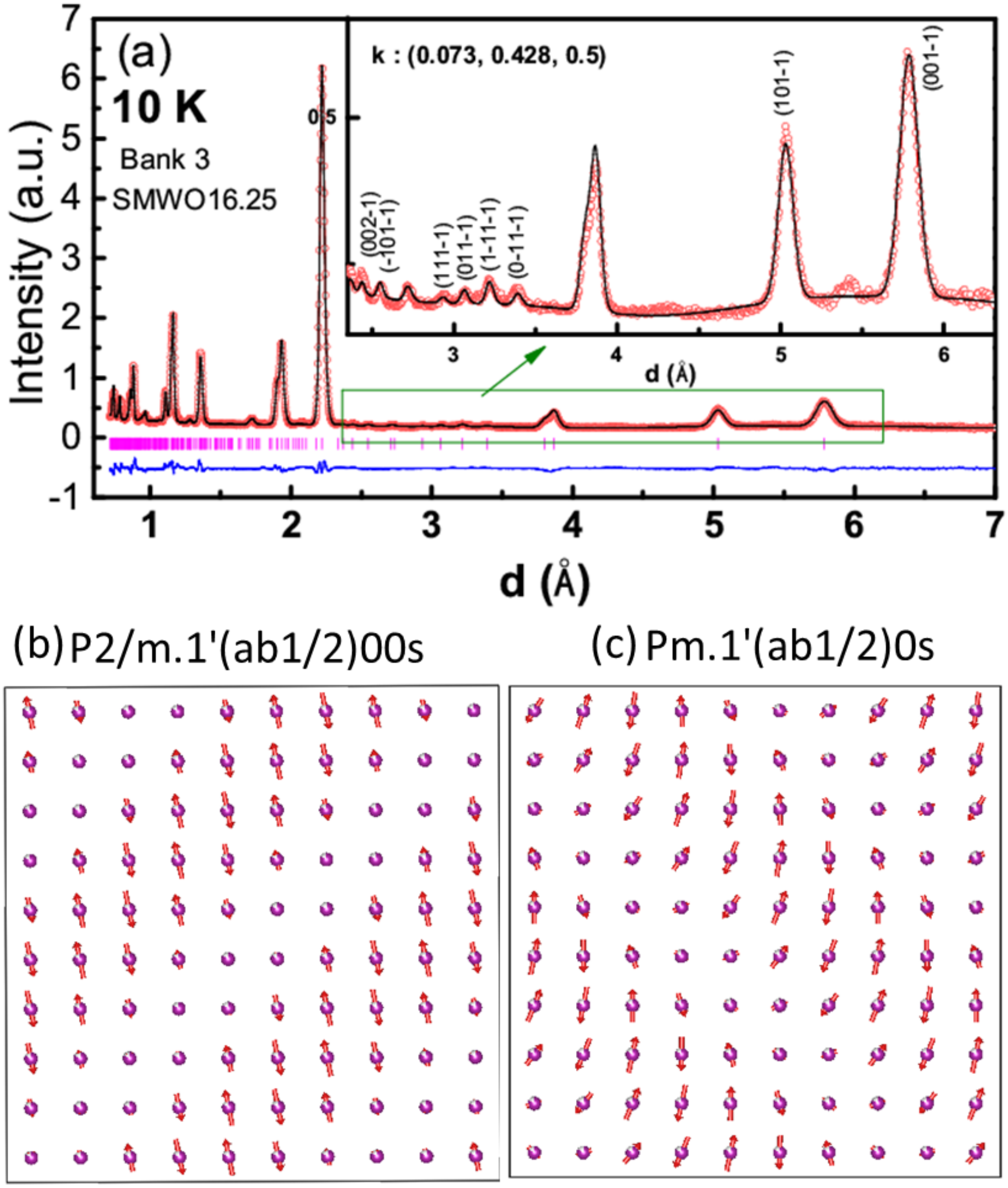

**Figure 7**

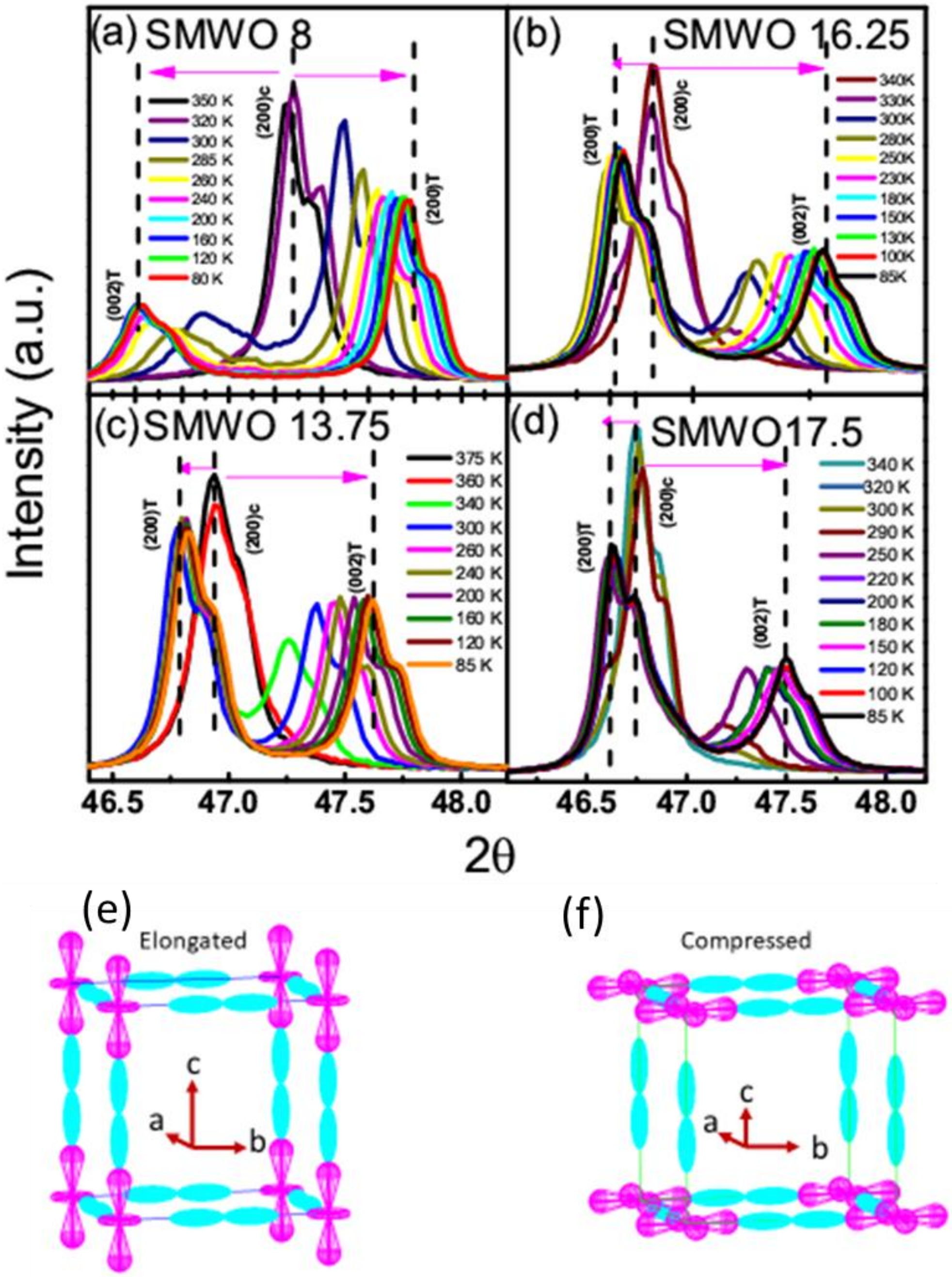

**Figure 8**

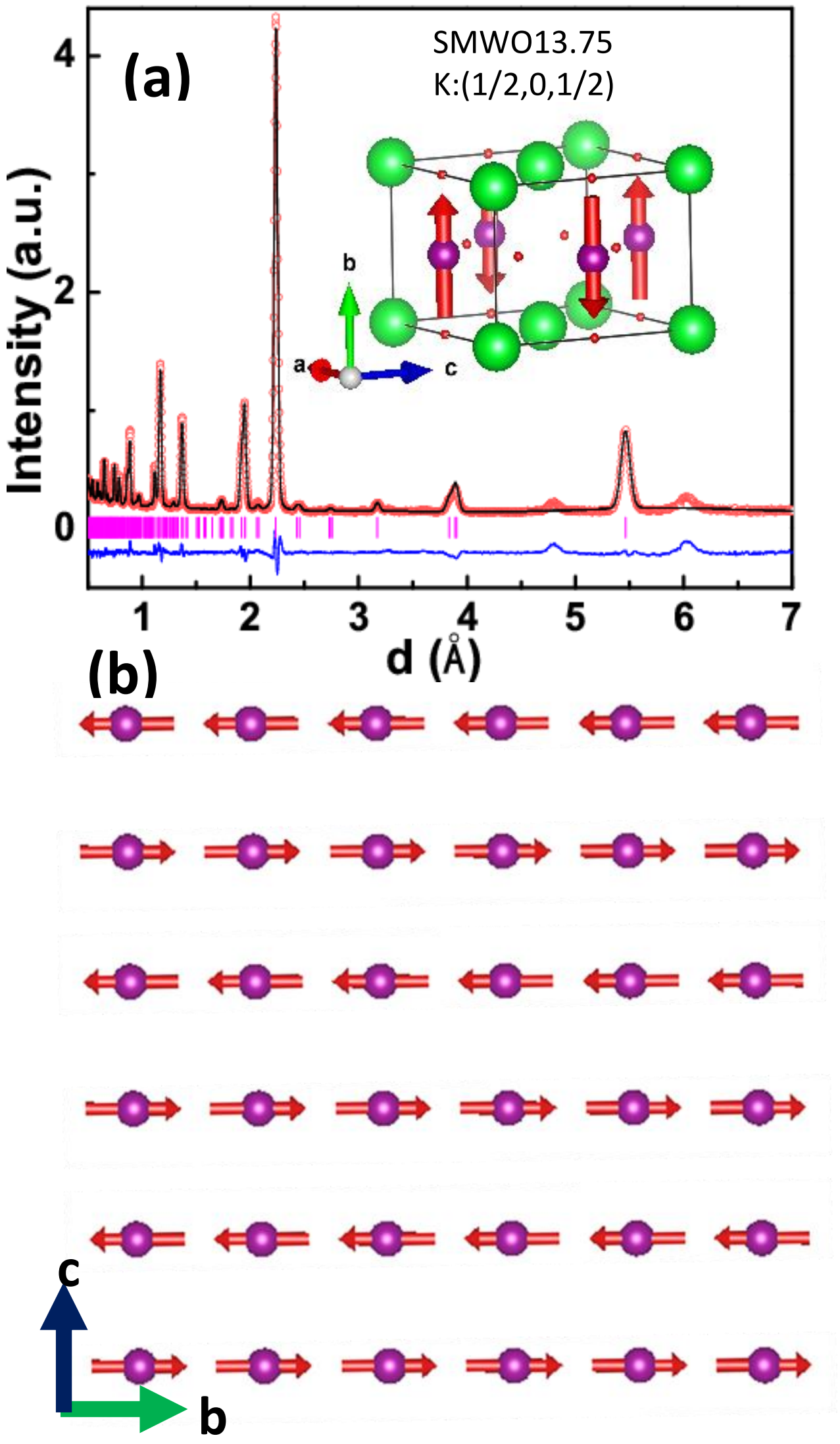

**Figure 9**

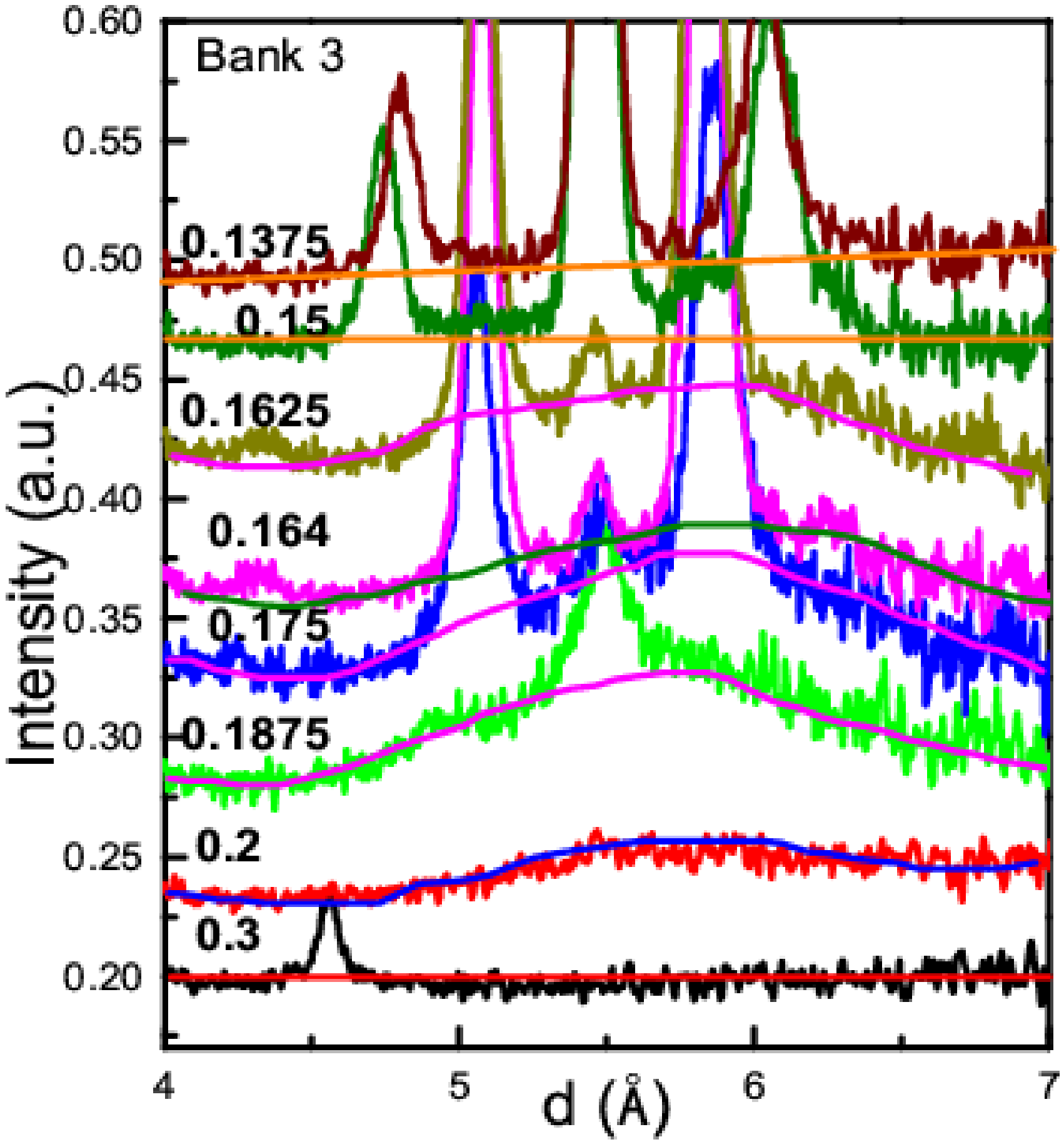

**Figure 10**

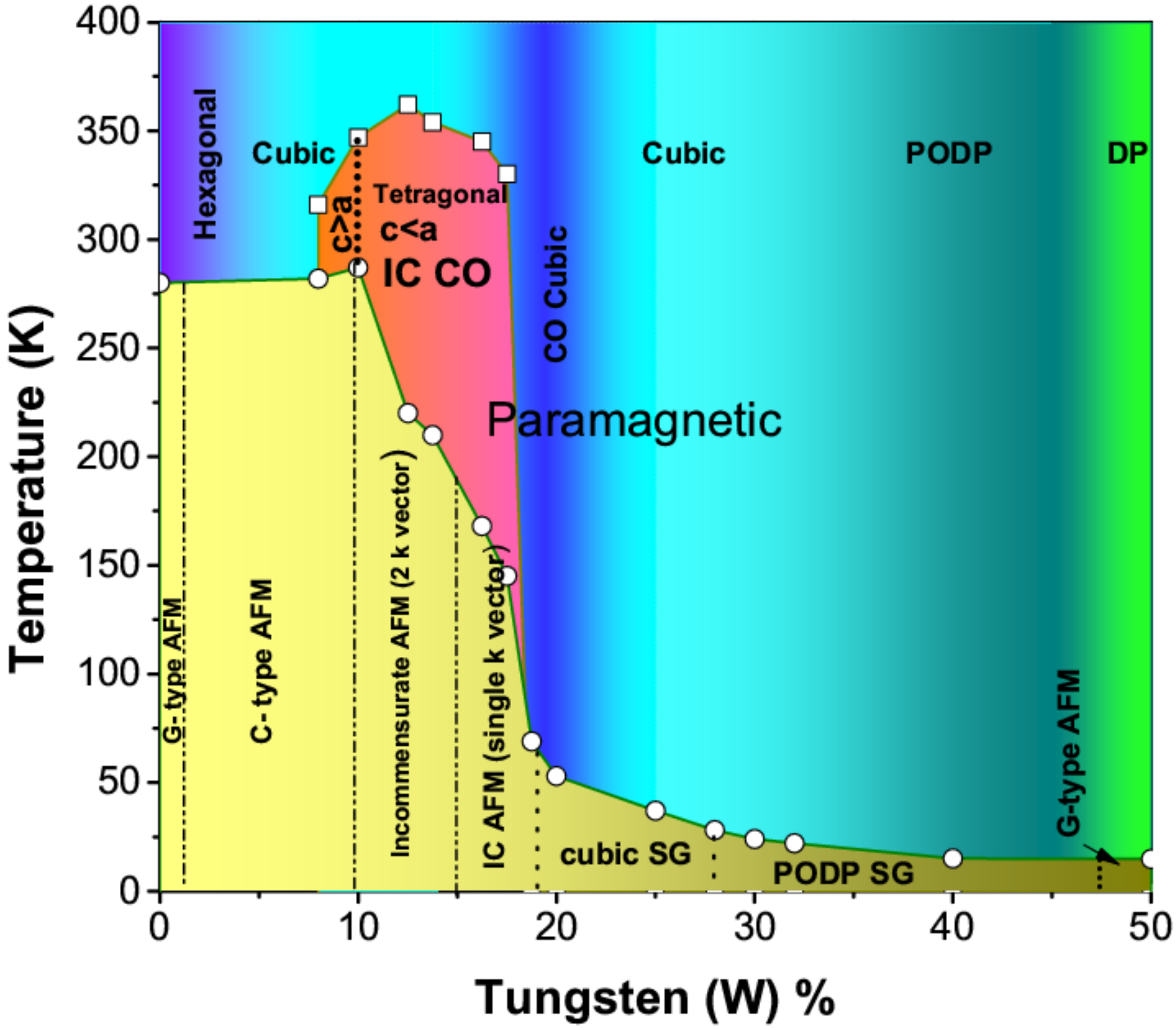

**Figure 11**